\documentclass[twocolumn,amsmath,amssymb]{revtex4}
\usepackage[dvips]{graphicx}
\usepackage{dcolumn}
\usepackage{bm}

\begin{document}
\title[]
{An unexpected topological censor}
\author{Peter K.\,F. Kuhfittig}
\email{kuhfitti@msoe.edu}
\address{Department of Mathematics\\
Milwaukee School of Engineering\\
Milwaukee, Wisconsin 53202-3109 USA}

\begin{abstract}\noindent
Morris-Thorne wormholes with a cosmological constant $\Lambda$ have been
studied extensively, even allowing $\Lambda$ to be replaced by a space
variable scalar. These wormholes cannot exist, however, if $\Lambda$ is
both space and time dependent. Such a $\Lambda$ will therefore
act as a topological censor.  While not likely to have a bearing
on the present, possible cosmological consequences of inflation
cannot be discounted. \\
\phantom{a}\\
PAC numbers: 04.20.Jb, 04.20.Gz
\end{abstract}

\maketitle
\noindent

\section{Introduction}\noindent
Wormholes are handles or tunnels in the spacetime topology connecting
two separate and distinct regions of spacetime.  These regions may be
part of our Universe or of different universes.  The pioneer work of
Morris and Thorne \cite{MT88} has shown that macroscopic wormholes may
be actual physical objects, provided that certain energy conditions
are violated.  Several wormhole studies have added the cosmological
constant $\Lambda$ \cite{DM95, DD01, jL03}.

When Einstein first introduced the cosmological constant into his
field equations in 1917, he was still striving for consistency with
Mach's principle.  From the standpoint of cosmology, however, $\Lambda$
served to create a kind of repulsive pressure to yield a stationary
Universe.  Eventually Zel'dovich identified $\Lambda$ with the vacuum
energy density due to quantum fluctuations \cite{yZ68}.

It has been proposed from time to time that the ``constant" is
actually a variable parameter.  For example, in discussing a
family of asymptotically flat globally regular solutions to the
Einstein field equations, Dymnikova \cite{iD02} notes that the
source term corresponds to an $r$-dependent $\Lambda$.  Assuming
that $\Lambda$ does indeed have the form $\Lambda=\Lambda(r)$,
Rahaman, \emph{et al.}, \cite{fR07} obtained a class of wormhole
solutions, while Ray, \emph{et al.}, \cite{RBS03} studied various
models that can be applied to the classical electron of the Lorentz
type.  Cosmic strings with $\Lambda=\Lambda(r)$ are discussed in Ref.
\cite{RRM06}.  In Ref. \cite{LW01} the variable $\Lambda$ is derived
from a higher spatial dimension and manifests itself as an
energy-density for the vacuum.

Another widely discussed possibility is a space- and time-dependent
$\Lambda$, i.e., $\Lambda=\Lambda(r,t)$, suggested by recent
observations of high redshift Type Ia supernovae
\cite{aR98, sP99, PR00, SS06, yM08}.  For a detailed discussion
with an extensive list of references, see Alcaniz \cite{jA06}.
For various $\Lambda$-decay scenarios from the original high value
during inflation to the present, see Ref. \cite{BS97} and references
therein.  Ref. \cite{jO99} discusses the big bang, as well as the
``big bounce," referring to variable-$\Lambda$ models having a
non-singular origin.

Using a natural extension of a metric proposed by Delgaty and
Mann \cite {DM95}, it is shown in this paper that if $\Lambda$ is
both space and time dependent, so that $\Lambda=\Lambda(r,t)$,
then a wormhole of the Morris-Thorne type will have a curvature
singularity at the center.  Possible cosmological implications
are discussed at the end.


\section{Background}\label{S:background}
\noindent
Using units in which $c=G=1,$ our starting point is the
Einstein-de Sitter metric
\begin{multline}\label{E:line1}
   ds^2=-\left(1-\frac{2M}{r}-\frac{\Lambda r^2}{3}\right)dt^2
   +\frac{dr^2}{1-\frac{2M}{r}-\frac{\Lambda r^2}{3}}\\
   +r^2(d\theta^2+\text{sin}^2\theta\,d\phi^2),
\end{multline}
which is the unique solution of the vacuum Einstein field equations
for a spherically symmetric spacetime with a positive cosmological
constant.  The line element reduces to the Schwarzschild line
element if $\Lambda=0.$  The wormhole metric in Ref.~\cite{MT88},
\begin{equation}\label{E:line2}
 ds^2=-e^{2\Phi(r)}dt^2+\frac{dr^2}{1-\frac{b(r)}{r}}+r^2(d\theta^2
     +\text{sin}^2\theta\,d\phi^2),
\end{equation}
provides a motivation for the following metric, proposed by
Delgaty and Mann \cite {DM95}:
\begin{multline}\label{E:line3}
   ds^2=-e^{2\Phi(r)}dt^2+\frac{dr^2}{1-\frac{M(r)}{r}-
   \frac{\Lambda(r,t)r^2}{3}}\\
   +r^2(d\theta^2+\text{sin}^2\theta\,d\phi^2).
\end{multline}
In Ref. \cite{DM95}, $\Lambda$ is fixed, while the constant
$1$ is incorporated in the function $M(r)$.  This metric
describes a traversable wormole in $(3+1)$ dimensions with
a cosmological constant $\Lambda$.    If $\Lambda$ is to
have the form $\Lambda=\Lambda(r,t)$, then Eq. (\ref{E:line3})
becomes the only natural choice for the new metric.

In the metric, Eq. (\ref{E:line3}), $\Phi(r)$ is called the
\emph{redshift function}.  If $\Lambda=0$, then $M(r)=b(r)$.  So
$M(r)$ will be called the \emph{shape function}; thus $M(r_0)=r_0$.
(Recall that in Eq.~(\ref{E:line2}), the sphere of radius $r=r_0$
is the \emph{throat} of the wormhole.)  Qualitatively, $M(r)$ has
the form shown in Fig. 1.  Observe that $\Lambda$ is a
\emph{positive} function of both $r$ and $t$.

According to Ref.~\cite{HV98}, since the wormhole described by the
metric in Eq.~(\ref{E:line3}) is dynamic, there are actually two
throats on opposite sides of the \emph{center} $r=r_1$.  This
center is determined implicitly (for any fixed $t$) from the
equation
\begin{equation}\label{E:throat}
     1-\frac{M(r)}{r}-\frac{\Lambda(r,t)r^2}{3}=0.
\end{equation}
(Observe that the entire sphere $r=r_1$ lies in the same time
slice.)  After rearranging terms, we get for any fixed
time-slice
\begin{equation}\label{E:fixpoint}
  F(r)=M(r)+\frac{1}{3}r^3\Lambda(r,t)=r.
\end{equation}
So for any fixed $t$, a solution to Eq.~(\ref{E:fixpoint}) is a fixed
point $F(r_1)=r_1.$
(See Fig. 1.) Since $M$, $\Lambda$, and $r$ are all positive,
\begin{figure}[htbp]
\begin{center}
\includegraphics [clip=true, draft=false, bb= 0 0 305 190,
angle=0, width=4.5 in,
height=2.5 in, viewport=40 40 302 185]{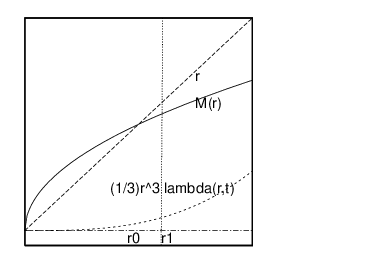}
\end{center}
\caption{\label{fig:figure1}Graph showing the fixed point $r=r_1$
of $F(r)$.}
\end{figure}
$M(r_1)<r_1$.  So $r_1>r_0$.  Since the sphere $r=r_1$ is the center
of the wormhole, $r=r_0$ is not in the manifold, while each throat is
a sphere with time-dependent radius $r_2>r_1$.

\section{The failed solution}\noindent
To study the presumptive wormhole solution, it is necessary to
compute the components of the Riemann curvature and Einstein tensors
using the following orthonormal basis:
\begin{align*}
  \theta^0&=e^{\phi(r)}dt, &\theta^1&=\left(1-\frac{M(r)}{r}-
    \frac{\Lambda(r,t)r^2}{3}\right)^{-1/2}dr,\\
   \theta^2&=r\,d\theta, &\theta^3&=r\,\text{sin}\,\theta\,d\phi.
\end{align*}
Some of the components of the Einstein tensor are listed next:
\begin{equation}\label{E:Einstein1}
  G_{\hat{t}\hat{t}}=\frac{M'(r)}{r^2}+\Lambda(r,t)+
   \frac{1}{3}r\frac{\partial}{\partial r}\Lambda(r,t),
\end{equation}
\begin{multline}\label{E:Einstein2}
   G_{\hat{r}\hat{r}}=\frac{2}{r}\left(1-\frac{M(r)}{r}-\frac
   {\Lambda(r,t)r^2}{3}\right)\Phi'(r)\\
     -\frac{M(r)}{r^3}-\frac{\Lambda(r,t)}{3},
\end{multline}
\begin{multline}\label{E:Einstein3}
   G_{\hat{r}\hat{t}}=\frac{r}{3}e^{-\Phi(r)}
       \frac{\partial}{\partial t}\Lambda(r,t)\times\\
    \left(1-\frac{M(r)}{r}-\frac{\Lambda(r,t)r^2}{3}\right)^{-1/2}.
\end{multline}
From the Einstein field equations with cosmological constant,
\begin{equation}
    G_{\hat{\alpha}\hat{\beta}}+\Lambda g_{\hat{\alpha}\hat{\beta}}
      =8\pi T_{\hat{\alpha}\hat{\beta}},
\end{equation}
we obtain
\begin{equation}
  T_{\hat{\alpha}\hat{\beta}}=\frac{1}{8\pi}
(G_{\hat{\alpha}\hat{\beta}}+\Lambda g_{\hat{\alpha}\hat{\beta}}).
\end{equation}
So
\begin{equation}\label{E:T1}
  T_{\hat{t}\hat{t}}=\rho(r,t)=\frac{1}{8\pi}\left(\frac{M'(r)}
  {r^2}+\frac{1}{3}r\frac{\partial}{\partial r}\Lambda(r,t)
     \right),
\end{equation}
\begin{multline}\label{E:T2}
   T_{\hat{r}\hat{r}}=p(r,t)=\frac{1}{8\pi}\left[\frac{2}{r}
   \left(1-\frac{M(r)}{r}-\frac{\Lambda(r,t)r^2}{3}\right)
   \Phi'(r)\right. \\
     \left.-\frac{M(r)}{r^3}+\frac{2}{3}\Lambda(r,t)\right],
\end{multline}
\begin{multline}\label{E:T3}
   T_{\hat{r}\hat{t}}=T_{\hat{t}\hat{r}}=-f(r,t)=\frac{1}{8\pi}
    \frac{r}{3}e^{-\Phi(r)}\frac{\partial}{\partial t}\Lambda(r,t)
     \times\\\\
     \left(1-\frac{M(r)}{r}-\frac{\Lambda(r,t)r^2}{3}\right)
     ^{-1/2},
\end{multline}
where $f(r,t)$ is usually interpreted as the energy flux in the
outward radial direction \cite{tR93}.

Now let us assume that at the throat $(r=r_2)$ the usual flare-out
conditions have been met and that for every $t$ the weak energy
condition (WEC) has been violated.  (The WEC states that  given the
stress-energy tensor $T_{\hat{\alpha}\hat{\beta}}$, the inequality
$T_{\hat{\alpha}\hat{\beta}}\mu^{\hat{\alpha}}\mu^{\hat{\beta}}
\ge 0$ holds for all time-like vectors and, by continuity, all null
vectors.) So for the  radial outgoing null vector $(1,1,0,0)$
we therefore have
\begin{equation}\label{E:WEC}
   T_{\hat{\alpha}\hat{\beta}}\mu^{\hat{\alpha}}\mu^{\hat{\beta}}
   =\rho+p\pm 2f<0.
\end{equation}
In this manner all the conditions for the existence of a wormhole
appear to have been met.  However, the real problem does not depend
on any violation of the WEC: in view of Eq.~(\ref{E:throat}),
we have for any given $t$
\begin{equation*}
  1-\frac{M(r_1)}{r_1}-\frac{\Lambda(r_1,t)r_1^2}{3}=0
\end{equation*}
at the center $r=r_1$.  Hence $f(r,t)$ cannot be a finite quantity
as long as $\partial\Lambda(r,t)/\partial t\ne 0.$  Similarly, the
components $G_{\hat{\theta}\hat{\theta}}$ and
$G_{\hat{\phi}\hat{\phi}}$,which are proportional to the lateral
pressure $p_t$, cannot be finite as long as $\Lambda$ is time
dependent:
\begin{multline}\label{E:G}
   G_{\hat{\theta}\hat{\theta}}=G_{\hat{\phi}\hat{\phi}}=
     -e^{-2\Phi(r)}\times \\
     \left [\frac{r^2}{6}\frac{\partial^2}{\partial t^2}
    \Lambda(r,t)\left(1-\frac{M(r)}{r}-\frac{\Lambda(r,t)r^2}{3}
        \right)^{-1}\right .\\
   \left .+\frac{r^4}{12}\left(\frac{\partial}{\partial t}\Lambda(r,t)
          \right)^2
   \left(1-\frac{M(r)}{r}-\frac{\Lambda(r,t)r^2}{3}\right)^{-2}
               \right ]\\
       +\text{other tems}.
\end{multline}

Finally, it is shown in Ref.~\cite{MT88} that for a wormhole to be
traversable by humanoid travelers, the radial tidal constraint must
be met: $|R_{\hat{t}\hat{r}\hat{r}}^{\phantom{\hat{t}\hat{r}\hat{r}}
\hat{t}}|\le(10^8\,\text{m})^{-2}$, where
$R_{\hat{t}\hat{r}\hat{r}}^{\phantom{\hat{t}\hat{r}\hat{r}}
\hat{t}}$ is a component of the Riemann curvature tensor.  This
component is given by
\begin{multline}\label{E:Riemann}
    R_{\hat{t}\hat{r}\hat{r}}^{\phantom{\hat{t}\hat{r}\hat{r}}
     \hat{t}}=-e^{-2\Phi(r)}\times\\
     \left[\frac{r^2}{6}\frac{\partial^2}{\partial t^2}
        \Lambda(r,t)
        \left(1-\frac{M(r)}{r}-\frac{\Lambda(r,t)r^2}{3}\right)
            ^{-1}\right.\\
        \left. +\frac{r^4}{12}\left(\frac{\partial}{\partial t}
         \Lambda(r,t) \right)^2
        \left(1-\frac{M(r)}{r}-\frac{\Lambda(r,t)r^2}{3}\right)
            ^{-2}\right]\\
        +\left(1-\frac{M(r)}{r}-\frac{\Lambda(r,t)r^2}{3}\right)
        \left(\Phi''(r)+[\Phi'(r)]^2\right)\\
      -\frac{1}{2}\Phi'(r)\left(\frac{M'(r)}{r}-
     \frac{M(r)}{r^2}\right.\\
     \left.+\frac{2}{3}r\Lambda(r,t)
           +\frac{1}{3}r^2\frac{\partial}{\partial r}
                \Lambda(r,t)\right).
\end{multline}
Because of Eq.~(\ref{E:throat}), we see that, once again, the right-hand
side of Eq.~(\ref{E:Riemann}) cannot be finite at the center as long as
$\Lambda$ is time dependent.  The same problem arises with the lateral
tidal constraints.  So even if the earlier problems did not occur, the
wormhole would not be traversable.

\section{A divergent scalar quantity}
\noindent
The singularities encountered so far could conceivably be removed
by a suitable coordinate transformation, as, for example, in the
Schwarzschild case.  To show that the spacetime is singular,
we need a scalar quantity that becomes infinite.  To this end we
list the components of the Ricci tensor.  First we define the function
\begin{multline*}
  H(r)=\left(1-\frac{M(r)}{r}-\frac{\Lambda(r,t)r^2}{3}\right)
   \left(-\Phi''(r)-[\Phi'(r)]^2\right)\\+\frac{1}{2}\Phi'(r)
   \left(\frac{rM'(r)-M(r)}{r^2}\right.\\+\left.
        \frac{2}{3}r\Lambda(r,t)
    +\frac{1}{3}r^2\frac{\partial}{\partial r}\Lambda(r,t)\right).
\end{multline*}
Then
\begin{equation*}
  R_{\hat{t}\hat{t}}=-H(r)+\frac{2}{r}
    \left(1-\frac{M(r)}{r}-\frac{\Lambda(r,t)r^2}{3}\right)
     \Phi'(r),
\end{equation*}
\vspace{0pt}
\begin{multline*}
  R_{\hat{r}\hat{r}}=H(r)+\frac{1}{r}\left(\frac{rM'(r)-M(r)}{r^2}
   \right.\\+\left.\frac{2}{3}r\Lambda(r,t)
    +\frac{1}{3}r^2\frac{\partial}{\partial r}\Lambda(r,t)\right),
\end{multline*}
\begin{multline*}
   R_{\hat{\theta}\hat{\theta}}=R_{\hat{\phi}\hat{\phi}}=-\frac{1}{r}
   \left(1-\frac{M(r)}{r}-\frac{\Lambda(r,t)r^2}{3}\right)
     \Phi'(r)\\+\frac{1}{2r}\left(\frac{rM'(r)-M(r)}{r^2}+
        \frac{2}{3}r\Lambda(r,t)
    +\frac{1}{3}r^2\frac{\partial}{\partial r}\Lambda(r,t)\right)\\
  +\frac{1}{r^2}\left(\frac{M(r)}{r}+\frac{\Lambda(r,t)r^2}{3}\right),
\end{multline*}
and
\begin{multline*}
  R_{\hat{r}\hat{t}}=\frac{r}{3}e^{-\Phi(r)}\frac{\partial}
   {\partial t}\Lambda(r,t)\left(1-\frac{M(r)}{r}
   -\frac{\Lambda(r,t)r^2}{3}\right)^{-1/2}.
\end{multline*}

Now consider the square of the curvature scalar
\begin{multline*}
  R_{\hat{\alpha}\hat{\beta}}R^{\hat{\alpha}\hat{\beta}}=\\
    R_{\hat{t}\hat{t}}R^{\hat{t}\hat{t}}
   +2R_{\hat{r}\hat{t}}R^{\hat{r}\hat{t}}
    +R_{\hat{r}\hat{r}}R^{\hat{r}\hat{r}}
   +R_{\hat{\theta}\hat{\theta}}R^{\hat{\theta}\hat{\theta}}
   +R_{\hat{\phi}\hat{\phi}}R^{\hat{\phi}\hat{\phi}}.
\end{multline*}
For any
fixed $t$ (that is, for any fixed time-slice), the term
$R_{\hat{r}\hat{t}}R^{\hat{r}\hat{t}}$ is divergent for some $r=r_1$
whenever
\[
    \frac{\partial}{\partial t}\Lambda(r,t)\ne 0.
\]
Being a scalar quantity, it diverges at the center in all coordinate
systems.

\section{Discussion}
\noindent
Before discussing the various implications, let us first recall
that the assumption
\[
    \frac{\partial}{\partial t}\Lambda(r,t)\ne 0
\]
has some clearcut consequences: Eqs. (\ref{E:T3}), (\ref{E:G}), and
(\ref{E:Riemann}) imply that the energy flux, lateral pressure,
and curvature cannot be finite at the center of the wormhole.
Since the scalar quantity $R_{\hat{\alpha}\hat{\beta}}
R^{\hat{\alpha}\hat{\beta}}$ also diverges, there is a curvature
singularity at the center.  So given the ansatz, Eq. (\ref{E:line3}),
it follows that for a wormhole of the Morris-Thorne type
to exist, $\Lambda$ must not be time dependent.  More formally,
using the language of the
topological censorship principle \cite{FSW93, CGS08},  causal curves
originating from and ending in a simply connected asymptotic region do
not see any non-trivial topology and can therefore be deformed to a
curve contained entirely within the asymptotic region.  In the present
situation, an ingoing radial null geodesic continues to move inward
and so cannot pass through the wormhole and probe the topology.  A
time-dependent $\Lambda$ will therefore act as a topological censor
for wormholes of the Morris-Thorne type.

Returning to the line element (\ref{E:line3}), suppose
$(\partial/\partial t)\Lambda(r,t)=0$ for $t\le t_0$ (for some
$t_0$) and that a Morris-Thorne wormhole exists.  If
$(\partial/\partial t)\Lambda(r,t)$ becomes nonzero for $t>t_0$,
then the center develops a curvature singularity.  So the entire
model, Eq. (\ref{E:line3}), breaks down and we no longer have
a valid wormhole solution.  With the properties of black holes
in mind, this singularity is not likely to disappear even if
$\Lambda$ becomes constant again.  Moreover,
since all the points on the sphere are singularities, the
infinite gravitational forces between them would pull the
entire sphere into a single point, thereby producing a
black hole.

While all these conclusions are based on fairly straightforward
calculations, one can question their relevance: for if $\Lambda$
really does change, then the rate of change is likely to be so
minute as to be practically undetectable.  Putting it another way,
even if $\Lambda$ should be independent of $r$, which is also
likely, $(\partial/\partial t)\Lambda(r,t)$ is going to be zero
within the margin of experimental error.  So the outcome has
no bearing on the present.

The situation would have been entirely different during a period
when $\Lambda$ really did change, at least with respect to time,
as would have been the case during inflation.  Here the existence
of a kind of \emph{vacuum energy} caused the Universe to act like
an approximation to a de Sitter solution since it was dominated
by a large effective cosmological ``constant" (Ref.
\cite{PR03}, p. 10).  At the very least, the change in $\Lambda$
would have been very large at the beginning of inflation, as well
as the end.  Now, submicroscopic wormholes existing prior to the
onset of inflation could conceivably have expanded to
macroscopic size \cite{tR93}.  However, such wormholes could not
have survived the beginning of inflation.

During inflation, $\Lambda$ would not only have been large, but
it may also have been constant.  (If not, wormholes could not
have formed.)  It is generally believed that inflation provides
a possible explanation for the initial inhomogeneities that have
led to the macroscopic structures we see today.  These
large-scale structures could have included wormholes.  But since
$\Lambda$ changed again rapidly at the end of inflation, such
wormholes could not have survived either.

These outcomes help explain why (apart from gravitational lensing)
the stars and galaxies observed are now believed to be distinct 
objects, rather than multiple images of a much smaller set.  Such 
multiple images would indeed
require a multiply-connected spacetime.  In addition,
the possibility that previously existing wormholes had become
black holes would help explain the large number of black holes
discovered, while the evidence for the existence of wormholes
is entirely lacking.

\end{document}